\documentclass[11pt,a4paper]{article}
\usepackage{a4wide}
\usepackage{epsfig}

\newcommand{\YFS}{$\Upsilon(4S)$}
\newcommand{\BF}[3]{{\cal B}(#1)=(#2)\times 10^{#3}}
\newcommand{\BFUL}[3]{{\cal B}(#1)<#2\times 10^{#3}}

\begin{document}
 \begin{flushright}
  KEK Preprint 2001-13\\
  KUNS-1717
 \end{flushright}
 \vspace*{1cm}
 \begin{center}
  \begin{Large}
   Radiative $B$ Meson Decay\\
  \end{Large}
  \vspace*{1cm}
  \begin{large}
   Yutaka Ushiroda\\
   ( \mbox{for the Belle Collaboration} )\\
  \end{large}
  \vspace*{0.5cm}
  Department of Physics, Kyoto University,\\
  Oiwake-cho Kitashirakawa Sakyo-ku, Kyoto, JAPAN
 \end{center}

 \vspace*{0.5cm}

 \begin{abstract}
  We have studied radiative $B$ meson decays using data collected at the
  \YFS\ resonance with the Belle detector at the KEKB $e^+\,e^-$
  storage ring.
  We measured the exclusive branching fractions to the $K^\ast(892)\gamma$
  final states to be
  $\BF{B^0\to {K^\ast}^0(892)\gamma}{49.6\pm 6.7\pm 4.5}{-6}$
  and
  $\BF{B^+\to {K^\ast}^+(892)\gamma}{38.9\pm 9.3\pm 4.1}{-6}$.
  The inclusive branching fraction is measured to be
  $\BF{B\to X_s\gamma}{3.39\pm 0.53\pm 0.42 {}^{+0.51}_{-0.55}}{-4}$.
  We searched for $B\to\rho\gamma$ decays and obtained an upper limit of
  ${\cal B}(B\to\rho\gamma)/{\cal B}(B\to K^\ast(892)\gamma)<0.19$ (90\% C.L.).
 \end{abstract}

 \vspace*{0.5cm}

 \begin{center}
  Contributed to the Proceedings of the 4th International Conference\\
  on $B$ Physics and $CP$ Violation, 19-23 February 2001, Ise-Shima, Japan
 \end{center}
 \vspace*{1cm}

 \section{Introduction}

 Radiative $B$ meson decays are sensitive to non-Standard-Model
 contributions.
 A Standard Model (SM) branching fraction calculation for $B\to
 X_s\gamma$ that includes next-to-leading order QCD corrections has a
 precision of 10\%~\cite{bib:SM}.
 Experimental measurements of the branching fraction are useful for
 identifying or limiting non-SM theories~\cite{bib:non-SM}.
 The $b{\to}d\gamma$ process may
 provide additional sensitivity to non-SM effects, a precise
 measurement of $|V_{td}/V_{ts}|$ and direct CP violation within the
 SM framework.  The exclusive channel $B\to\rho\gamma$ can be
 distinguished from the $B\to K^\ast(892)\gamma$ signal utilizing good
 high momentum kaon identification devices.

 We have analyzed a large data sample collected at the
 \YFS\ resonance with the Belle detector~\cite{bib:Belle} at the KEKB $e^+e^-$
 storage ring~\cite{bib:KEKB}.  We used 6.1 million $B\bar B$ events for $B\to
 X_s\gamma$ analysis and 11 million for other analyses.

 \section{Analyses and Results}
 We have performed measurements of branching fractions for
 $B\to K^\ast(892)\gamma$, $B\to K_2^\ast(1430)\gamma$ and $B\to
 X_s\gamma$, and a search for $B\to\rho\gamma$.
 We also measured the partial rate asymmetry in $B\to
 K^\ast(892)\gamma$.
 We will give the results with descriptions for the essential part of
 analyses in the following subsections. 

 \subsection{$B\to K^\ast(892)\gamma$}
 We did an exclusive reconstruction for $B\to K^\ast(892)\gamma$, where
 $K^\ast(892)$ is reconstructed from $K^\pm\pi^\mp$, $K^0_S\pi^0$,
 $K^0_S\pi^\pm$ and $K^\pm\pi^0$.
 A combined kaon-to-pion likelihood is constructed from the aerogel \v
 Cerenkov counters (ACC), the time-of-flight counter (TOF) and the
 $dE/dx$ from the central drift chamber (CDC).
 A $K/\pi$ separation cut is applied on the likelihood ratio.
 A $K\pi$ invariant mass window of $\pm 75\rm\,MeV/c^2$ is taken around 
 the nominal $K^\ast(892)$ mass.
 A likelihood ratio cut is applied in order to suppress the $q\bar q$
 background, where the likelihood ratio is calculated from the $B$ meson 
 flight direction, the helicity angle of the $K^\ast(892)$
 decay, and an event shape variable which we call {\it the Super
 Fox-Wolfram} (SFW).
 The SFW variable is a Fisher discriminant formed using Fox-Wolfram moments.
 The definition of SFW can be found in reference~\cite{bib:Xsgamma-Prepri}.
 
\begin{figure}
  \begin{center}
   \epsfig{figure=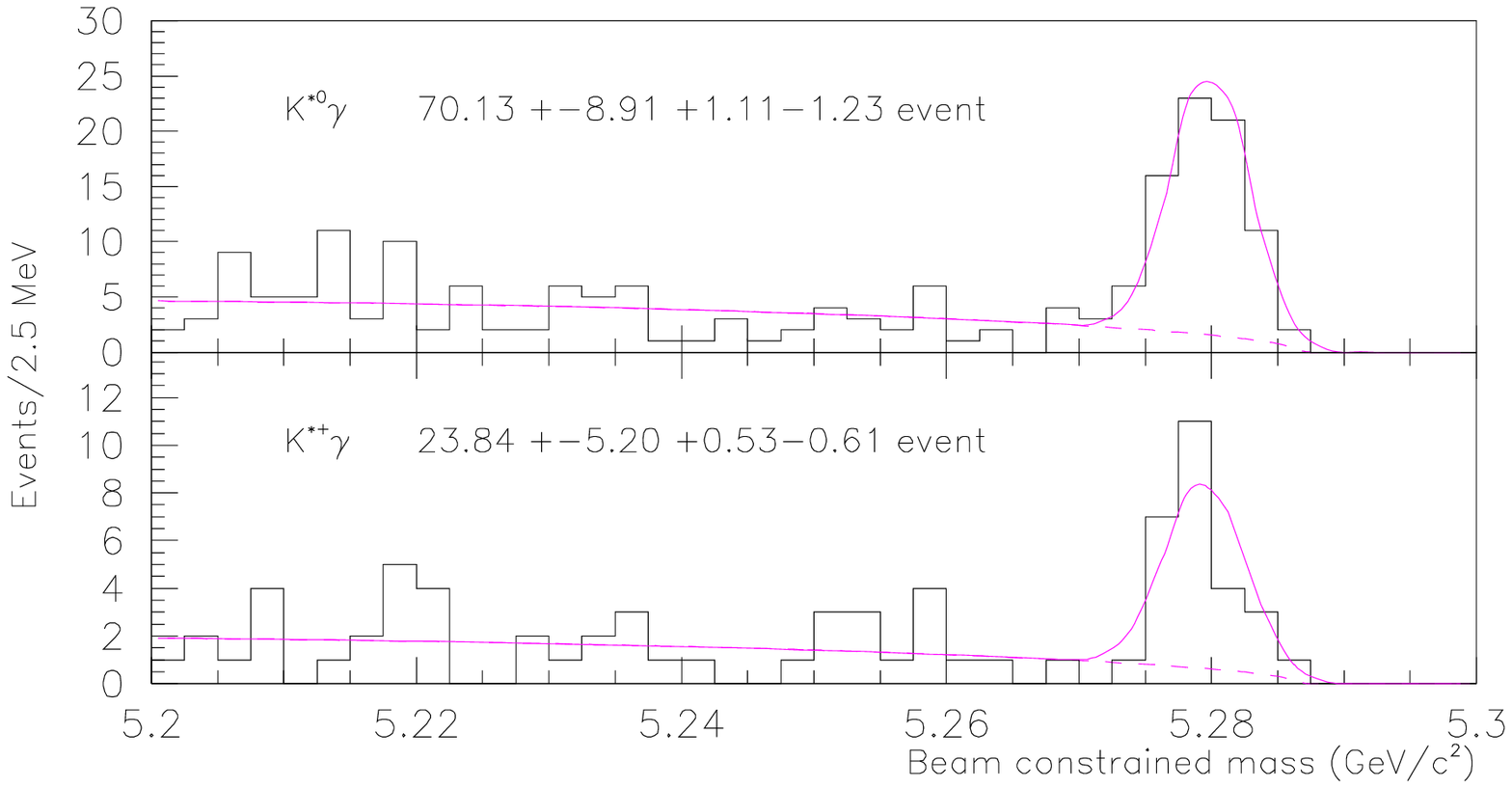,width=0.6\textwidth}
  \end{center} 
  \caption{Beam constrained mass spectra for $B\to K^\ast\gamma$ --
  ${K^\ast}^0\gamma$(top) and ${K^\ast}^+\gamma$(bottom).}
  \label{fig:mbkst892}
 \end{figure}

 We extracted the signal yields by fitting the beam constrained mass ($M_{bc}$) distribution
 with an ARGUS function for the background and a Gaussian for the signal 
 as shown in Figure~\ref{fig:mbkst892}.
 We determined the branching fractions to be
 \[
 \BF{B\to {K^\ast}^0(892)\gamma}{49.6\pm 6.7\pm 4.5}{-6}
 \]
 \[
 \BF{B\to {K^\ast}^+(892)\gamma}{38.9\pm 9.3\pm 4.1}{-6}. 
 \]

 We then checked the partial rate asymmetry,
 \[
  A_{CP} = \frac{1}{1-2\eta}\cdot
   \frac{{\cal N}(\bar K^{\ast 0}\gamma + K^{\ast -}\gamma)
        -{\cal N}(     K^{\ast 0}\gamma + K^{\ast +}\gamma)}
	{{\cal N}(\bar K^{\ast 0}\gamma + K^{\ast -}\gamma)
        +{\cal N}(     K^{\ast 0}\gamma + K^{\ast +}\gamma)}
 \]
 where ${\cal N}$ represents the yield and $\eta$ is the wrong tag fraction.
 $|A_{CP}|$ is expected to be less than 1\% in the SM.
 We use only {\it self-tagging} modes, namely we exclude the
 $K^0_S\pi^0\gamma$ mode.
 Thanks to Belle's good kaon identification devices, the wrong tag fraction
 $\eta$ is only 1.2\%.
 We have checked the intrinsic asymmetry in our analysis with an
 inclusive $K^\ast(892)$ sample, in which the momentum range of the
 $K^\ast(892)$ is adjusted to match our singal.  We found that the intrinsic
 asymmetry is $0.003\pm 0.01$, and is consistent with zero.
 We extracted the partial yields
 ${\cal N}(\bar K^{\ast 0}\gamma + K^{\ast -}\gamma) = 46.9\pm 7.3$
 and
 ${\cal N}(     K^{\ast 0}\gamma + K^{\ast +}\gamma) = 44.8\pm 7.1$.
 We calculated the partial rate asymmetry as
 \[
  A_{CP} = +0.02\pm 0.11\pm 0.01
 \]
 and found it is consistent with zero.

 \subsection{$B\to K_2^\ast(1430)\gamma$}
 We performed a similar analysis with a $K\pi$ invariant mass window of $\pm
 100\rm\,MeV/c^2$ around the nominal $K_2^\ast(1430)$ mass.
 We removed the helicity angle from the likelihood ratio in order to
 check the helicity distribution. This is because there is an
 overlapping $K^\ast(1410)$ resonance which has a different spin and thus
 a different helicity distribution.

 \begin{figure}
  \begin{center}
   \epsfig{figure=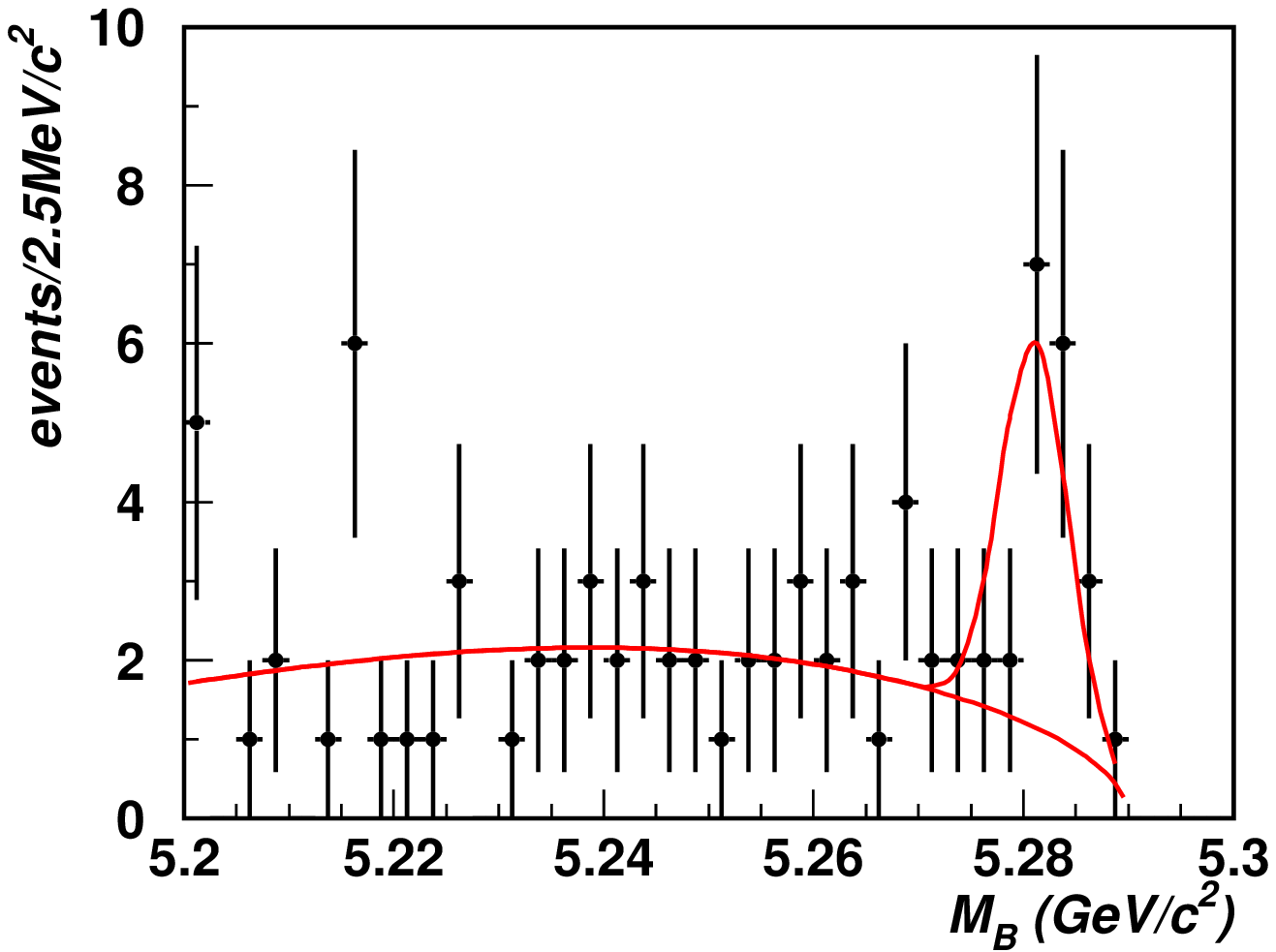,width=0.6\textwidth}
  \end{center}
  \caption{Beam constrained mass spectrum for $B\to ``K_2^\ast(1430)\mbox{''}\gamma$ candidates.}
  \label{fig:mbk2st}
 \end{figure}

\begin{figure}
 \begin{center}
  \epsfig{figure=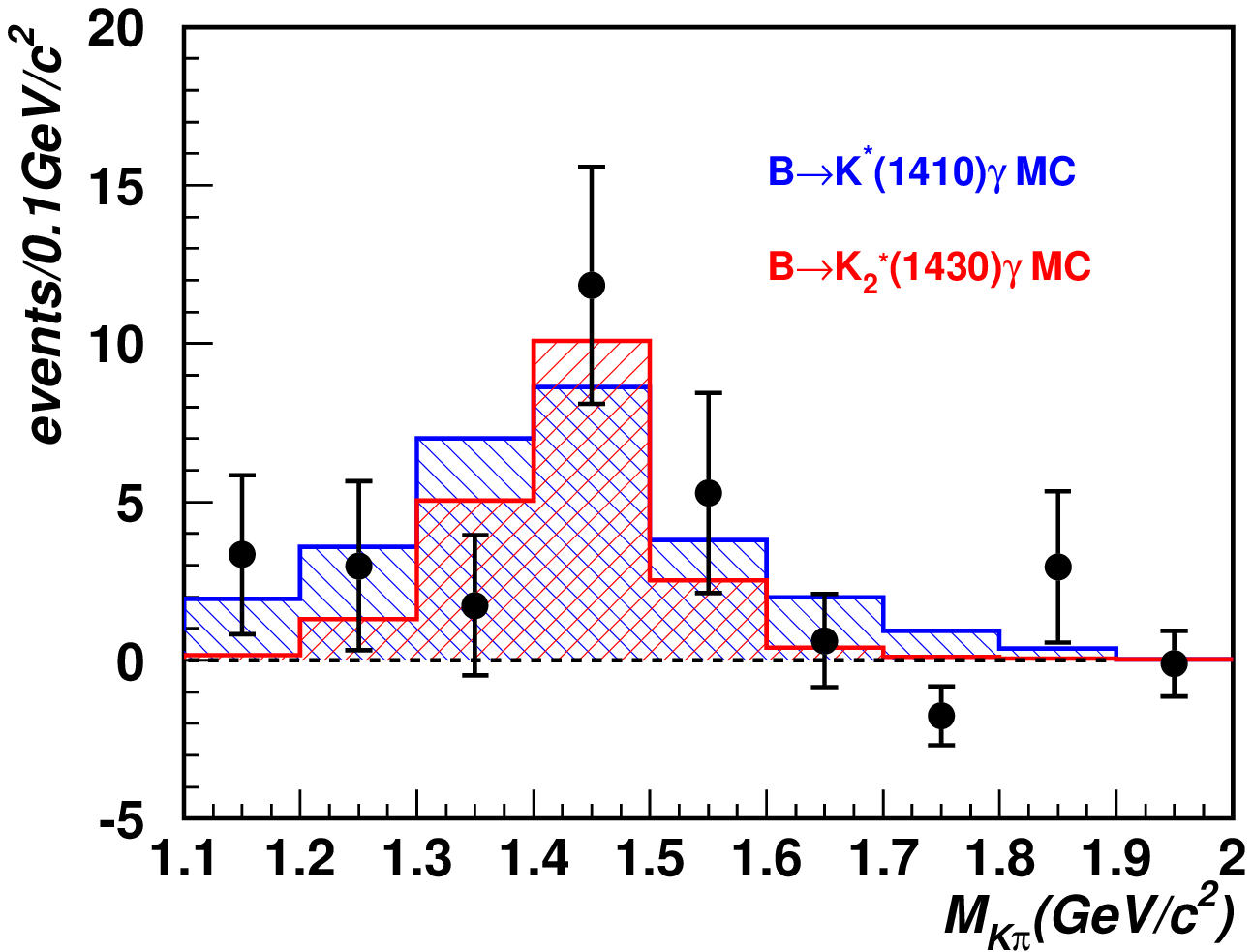,width=0.6\textwidth}
 \end{center}  
  \caption{$K\pi$ invariant mass spectrum for $B\to ``K_2^\ast(1430)\mbox{''}\gamma$ candidates.}
  \label{fig:mk2st}
 \end{figure}

\begin{figure}
  \begin{center}
   \epsfig{figure=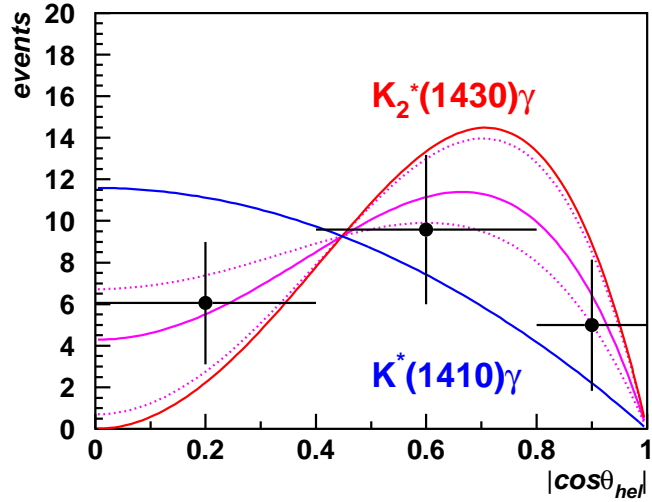,width=0.6\textwidth}
  \end{center} 
  \caption{Helicity distribution for $B\to
  ``K_2^\ast(1430)\mbox{''}\gamma$ candidates.  The result of the fit
  and the fitting functions $1-x^2$ and $x^2-x^4$ are shown as solid
  lines. Dotted lines are one standard deviation of the fit result.}
  \label{fig:helk2st}
 \end{figure}

 We observed $15.6\pm 4.6 {}^{+0.6}_{-0.7}$ events as shown in
 Figure~\ref{fig:mbk2st}.
 In Figure~\ref{fig:mk2st}, we see a resonance peak in the $K\pi$ invariant
 mass spectrum.
 We fitted the helicity distribution with curves for spin-1 and spin-2
 hypotheses as shown in Figure~\ref{fig:helk2st} and obtained
 the $K_2^\ast(1430)\gamma$ fraction to be ($63\pm 31$)\% where the error 
 is statistical only.
 Assuming these are all $K_2^\ast(1430)\gamma$ events, we quote the
 branching fraction to be
 \[
 \BF{B\to ``K_2^\ast(1430)\mbox{''}\gamma}{18.9\pm 5.6\pm 1.8}{-6}.  
 \]

 \subsection{$B\to\rho\gamma$}
 We combined $\pi\pi$ instead of $K\pi$ to search for $B\to\rho\gamma$.
 We took a mass window of $\pm 150\rm\,MeV/c^2$ around the nominal $\rho$
 mass. To suppress the $B\to K^\ast(892)\gamma$ contribution, we vetoed
 kaons by 
 a tight kaon-to-pion likelihood ratio cut, and vetoed the
 $\pi\pi$ combination if its invariant mass with a $K\pi$ mass hypothesis
 falls within a window of $50\rm\,MeV/c^2$ around the nominal $K^\ast(892)$ mass.
 The $K^\ast(892)\gamma$ background contribution is expected to be only
 0.4 (0.02) events for $\rho^0\gamma$ ($\rho^+\gamma$).

\begin{figure}
  \begin{center}
   \epsfig{figure=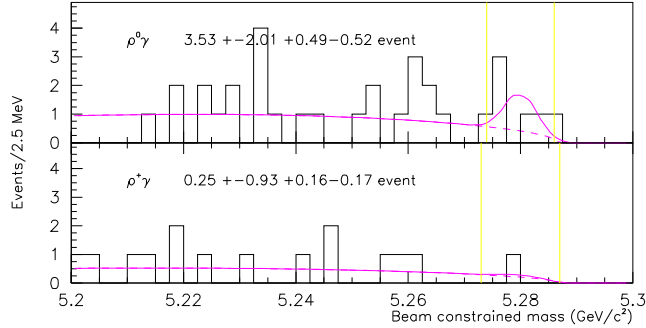,width=0.6\textwidth}
  \end{center} 
  \caption{Beam constrained mass spectrum searching for
  $B\to\rho^0\gamma$ (top) and $B\to\rho^+\gamma$(bottom).}
  \label{fig:mbrho}
 \end{figure}

 We observed 6 (1) candidates and a $3.5\pm 2.1$ ($0.3\pm 0.9$) event
 excess including $K^\ast(892)\gamma$ background for $B\to\rho^0\gamma$
 ($B\to\rho^+\gamma$) as shown in Figure~\ref{fig:mbrho}.
 Since neither is significant, we set
 upper limits for the branching fractions
 $\BFUL{B\to\rho^0\gamma}{10.6}{-6}$ and
 $\BFUL{B\to\rho^+\gamma}{9.9}{-6}$
 at 90\% confidence level.
 We also set an upper limit on the $\rho\gamma$ to
 $K^\ast(892)\gamma$ fraction,
 $\frac{{\cal B}(B\to\rho\gamma)}{{\cal B}(B\to K^\ast(892)\gamma)}<0.19$
 at 90\% confidence level. 

 \subsection{$B\to X_s\gamma$}
 A semi-inclusive reconstruction is performed by summing up multiple
 final states. We reconstruct the recoil $X_s$ system from
 one $K^\pm$ or $K^0_S$ and one to four pions of which at most one pion can
 be a $\pi^0$.  From $X_s$ and the most energetic photon candidate, we 
 form two independent kinematic variables, $M_{bc}$ and the energy
 difference $\Delta E$ in the \YFS\ rest frame.  We also calculate the
 angle between the $\gamma$ and the $X_s$ and require that they are
 back-to-back in the \YFS\ rest frame.

 We require an $X_s$ candidate to form a vertex with the beam profile
 constraint except in the case of the $K^0_S\pi^0\gamma$ mode which has no
 charged track to form a vertex. When there are multiple candidates in
 an event, we choose the one with the best vertex confidence level.  To
 solve the ambiguities due to neutral particles which do not affect
 the vertex, we choose the candidate for which the angle between the photon and
 the $X_s$ is the largest.

\begin{figure}
  \begin{center}
   \epsfig{figure=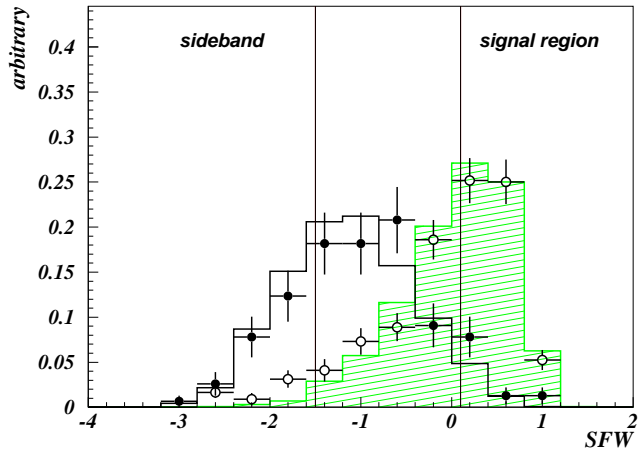,width=0.6\textwidth}
  \end{center}
  \caption{SFW distribution comparison between signal and $q\bar q$
  background -- The signal is concentrated on the higher side while the
  $q\bar q$ background is shifted to the lower side. Histograms are
  signal MC (hatched) and $q\bar q$ MC (open), error bars are $B\to
  D\pi$ data (open circles) and off-resonance data (solid circles).}
  \label{fig:sfw}
 \end{figure}

 For the selected single candidate,
 we require SFW to be greater than 0.1 to
 suppress the background from $q\bar q$ light quark pair production.
 We also require the recoil mass to be less than
 $2.05\rm\,GeV/c^2$ to suppress the background from other $B\bar B$
 decays. The SFW distributions for the signal and the $q\bar q$
 background are shown in Figure~\ref{fig:sfw}.

 In order to subtract the contribution from $q\bar q$ background, which is
 the main background in this analysis, we use the SFW
 sideband data (SFW$<-1.5$) instead of the statistically limited off-resonance data.
 The SFW variable is tailored not to be correlated with $M_{bc}$,
 and hence the $M_{bc}$ distributions in the signal region and in the
 sideband region are the same.

\begin{figure}
   \begin{center}
    \epsfig{figure=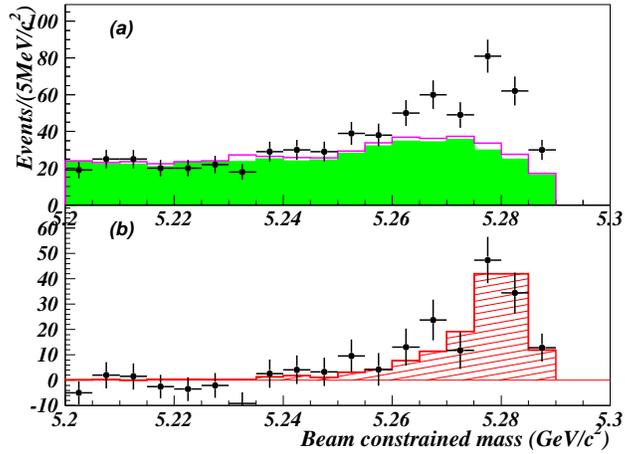,width=0.6\textwidth}
   \end{center}
   \caption{Beam constrained mass spectrum for $B\to X_s\gamma$ :
   (a) observed spectrum (points with error bars) and background contributions;
   the total background is the open histogram
   and the $q\bar q$ contribution is the solid part.
   (b) background subtracted data (points with error bars) and signal MC prediction
   (hatched histogram).}
  \label{fig:mbxs}
 \end{figure}

 We estimate the $B\bar B$ background by using Monte Carlo (MC) and subtract
 its contribution from the observed data.  We then fit the observed
 $M_{bc}$ spectrum (Figure~\ref{fig:mbxs}) to be sum of the sideband
 data and the signal MC floating the normalizations.
 We find a signal yield of $106.5\pm 16.8\pm 5.0$ events,
 where the first error is
 statistical and the second is the systematic error in our $q\bar q$
 background estimation.

\begin{figure}
   \begin{center}
    \epsfig{figure=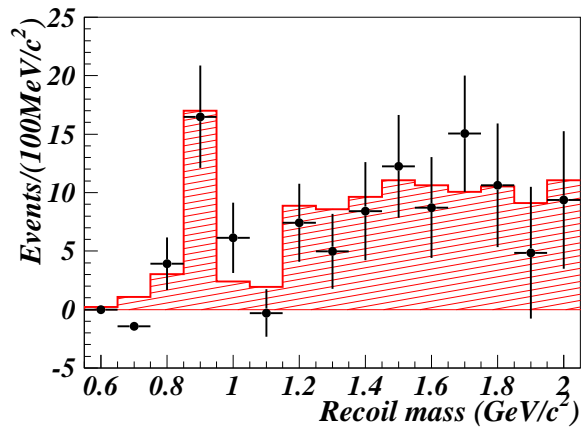,width=0.6\textwidth}
   \end{center} 
   \caption{Recoil mass spectrum for $B\to X_s\gamma$.
   Data  points (points with error bars) are compared with signal MC (hatched histogram).}
  \label{fig:mxs}
 \end{figure}

 In the recoil mass spectrum (Figure~\ref{fig:mxs}),
 we see a clear $K^\ast(892)$ mass
 peak and a continuum contribution in the higher resonance region.
 Therefore, we modeled the recoil mass system as a mixture of
 $K^\ast(892)$ and a continuum contribution with the observed ratio of
 the yields in the $K^\ast(892)$ region to the continuum region.
 For the continuum part of the recoil mass spectrum, we adopted a model proposed by
 Kagan and Neubert~\cite{bib:KaganNeubert} with the input $b$ quark mass
 parameter equal to $4.75\pm 0.10\rm\,GeV/c^2$.

 \begin{figure}
  \begin{center}
   \epsfig{figure=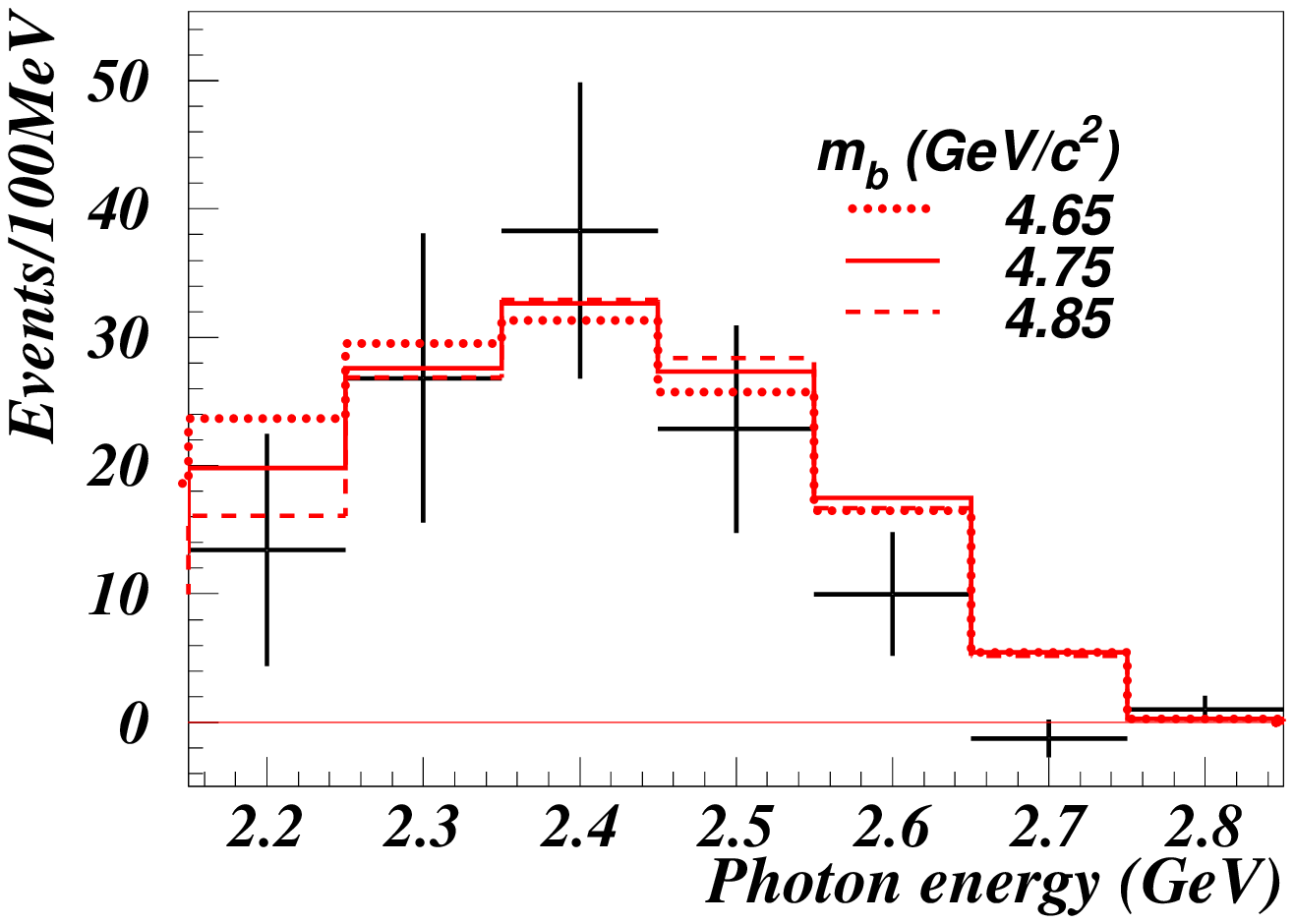,width=0.6\textwidth}
  \end{center} 
  \caption{Photon energy spectrum for $B\to X_s\gamma$.
  Data points (points with error bars) are compared to signal MC (histograms) with
  three different $b$ quark mass parameter values.}
  \label{fig:egamma}
 \end{figure}

 We determined the branching fraction
 \[
 \BF{B\to X_s\gamma}{3.36\pm 0.53\pm 0.42 {}^{+0.50}_{-0.54}}{-4}
 \]
 where the first error is statistical, the second is systematic and the
 third is the model error.
 In Figure~\ref{fig:egamma}, we show the measured photon energy spectrum 
 which is corrected for our recoil mass cut. This spectrum can be
 compared with the model.

 \section{Summary}
 The measured branching fractions for $B\to K^\ast(892)\gamma$
 and $B\to X_s\gamma$ are consistent with the other experimental results~\cite{bib:experiment}
 and/or the SM prediction~\cite{bib:SM}.  We do not see a deviation from the SM prediction 
 in the partial rate asymmetry in $B\to K^\ast(892)\gamma$, either.

 We observed a significant excess of $B\to (K\pi)\gamma$ in the
 $K_2^\ast(1430)$ resonance region.  The contributions from
 $K_2^\ast(1430)$ and $K^\ast(1410)$ are not disentangled yet.

 We have not observed $B\to\rho\gamma$ decays.  The upper limit for its
 branching fraction has been updated from what we reported
 previously~\cite{bib:ICHEP2000}.


  
\end{document}